# AES Encryption and Decryption Using Direct3D 10 API

Adrian Marius Chiuţă

*Computer Science Department, IT&C Security Master*
*Cybernetics and Economic Informatics*
*Academy of Economic Studies*
*Calea Dorobanţilor, Nb. 15-17, Room 2315, Sector 1, Bucharest*
*ROMANIA*
*achiuta@tessera.com, http://ism.ase.ro*

**Abstract:** Current video cards (GPUs – Graphics Processing Units) are very programmable, have become much more powerful than the CPUs and they are very affordable. In this paper, we present an implementation for the AES algorithm using Direct3D 10 certified GPUs. The graphics API Direct3D 10 is the first version that allows the use of integer operations, making from the traditional GPUs (that works only with floating point numbers), General Purpose GPUs that can be used for a large number of algorithms, including encryption. We present the performance of the symmetric key encryption algorithm – AES, on a middle range GPU and on a middle range quad core CPU. On the testing system, the developed solution is almost 3 times faster on the GPU than on one single core CPU, showing that the GPU can perform as an efficient cryptographic accelerator.

## 1. Introduction

Because of the importance given more increasingly to the information security for both, servers and personal computers, in some specific cases, from communications to active storage and back-up, a lot of processing time is used to encrypt or decrypt data. To not use all the processing power of CPUs only for encryption and decryption, specific accelerators have been developed solely for this purpose. Currently, one of the most powerful GPU (AMD Radeon 5870) can perform in the best case scenario 2.72 TFlops and has the memory bandwidth of 153.6 Gbytes/sec, and one of the most powerful desktop CPUs (Intel Core i7 965EE, 4 cores, 8 threads) has a processing power of about 55 GFlops and a memory bandwidth of 17 Gbytes/sec.

As shown, current GPUs have a tremendous processing power while the algorithm running on them is very parallelizable, and the purpose of this paper is to use these resources through the implementation of cryptographic functions on the GPU, so leaving the CPU free for other activities, and rendering useless the need for expensive cryptographic accelerators. Since most current systems have powerful graphics cards, but they are rarely used at full capacity, GPUs can be used as a cryptographic coprocessor or accelerator. The cryptographic processing can be performed partly or entirely on the GPU for different systems and applications, from servers to laptops, from applications used for communications, to applications that carry large volumes of cryptographic operations, such as backup and archiving applications.

In this paper we will present the implementation on GPUs of AES algorithm using Direct3D 10 API for Windows Vista.

## 2. Current problems and solutions

### 2.1. Block Cipher Algorithms and AES Details

Block cipher algorithms are symmetric key chippers that operate on a fixed number of bits (a block), having as input *n*-bits of plain text and as output *n*-bits of ciphered text. If the input plain text is longer than $n$-bits, it will be divided in blocks of $n$-bits and if necessary padded at the end with known values, each block being processed separately. The decryption is similar with the encryption, all the steps from the encryption being performed in inverse order. AES is the latest symmetric key algorithm, it was announced by National Institute of Standards and Technology (NIST) as U.S. FIPS PUB 197 [1] and it was declared as standard on 26 May 2002. AES is a simplified version of the Rijndael algorithm developed by Joan Daemen and Vincent Rijmen, and opposed to Rijndael algorithm it supports only blocks of 128 bits and key size of 128, 192 and 256 bits. The AES algorithm consists of a variable number of rounds, depending on the size of the key (10, 12 respective 14 rounds), and each round operate on a state matrix that consist of 4x4 bytes (the same size as the blocks). For encryption and also for decryption, AES algorithm uses a function composed of four round transformations at the byte level: *SubBytes* (byte-level substitution using a substitution table - S-Box), *ShiftRows* (rotation of lines for state matrix using different offsets), *MixColumns* (mixing of data in each column of the state matrix), and *AddRoundKey* (adding a key round in state matrix). All these transformations are applied on the state and are based on operations performed in a finite particular field called Galois - GF(pn). In this paper we use an optimized version of the AES algorithm, where the state is formed by four 32 bit values and the transformations *SubBytes*, *ShiftRows* and *MixColumns* are performed in the same step (the *S-Box* and the four tables used by *MixColumns* are combined into four tables $T_i$, $i<4$). Using these optimizations, one round can be written like:

(1)

$O_0 = T_0[I_{0,0}]$ ^ $T_1[I_{1,1}]$ ^ $T_2[I_{2,2}]$ ^ $T_3[I_{3,3}]$ ^ $K_0$

$O_1 = T_0[I_{1,0}]$ ^ $T_1[I_{2,1}]$ ^ $T_2[I_{3,2}]$ ^ $T_3[I_{0,3}]$ ^ $K_1$

$O_2 = T_0[I_{2,0}]$ ^ $T_1[I_{3,1}]$ ^ $T_2[I_{0,2}]$ ^ $T_3[I_{1,3}]$ ^ $K_2$

$O_3 = T_0[I_{3,0}]$ ^ $T_1[I_{0,1}]$ ^ $T_2[I_{1,2}]$ ^ $T_3[I_{2,3}]$ ^ $K_3$

where $T_i[]$ represent a memory access to one of the precomputed tables, $I_{i,k}$ represent the $k$-th byte of the $i$-th 32 bit input, $K_i$ represent the $i$-th 32 bit part of the round key, $O_i$ represent the $i$-th 32 bit output value from that round and ^ represent XOR. So basicly, one round for the AES algorithm consists at least from 16 memory accesses and 16 XOR operations, and we need to perform at least ten rounds for a block (for 128-bit key size), so for 128 bits of input data we will have to perform at least 160 memory accesses and 160 XOR operations. From this numbers we can see that the encryption using AES algorithm is a very time consuming operation that is why alternatives are seached to offload this from the CPU.

For more details about AES algorithm refer to [1].

## 2.2. Known Implementations on GPUs

One of the first implementations of AES algorithm on GPU is described in Cook et al. [2], and is using only the fixed functions from the OpenGL API. This approach is indeed working on a wide range of GPUs, but with very poor performance even for high end GPUs because CPU resources will be wasted on preprocessing the input data that is sent to the GPU (the input data must have a specific format). Using this method, the authors achieved an encryption speed of 0.191 Mbytes/sec for 128-bit ECB AES on GeForce 3 GPU (76 GFlops).

Another implementation using also the OpenGL API, but this time using *shaders* to program the GPU, is described in Harrison et al. [3]. In this paper, the authors have implemented two variants for the algorithm: first variant uses only GPU *shader* programming, but because at the time of the implementation were available only GPUs that worked only with floating point number, each used register contained an 8 bit value encoded as floating point, and for XOR operation look-up tables were used. The second variant also GPU *shader* programming, but XOR operations are performed using fixed functions.This variant is faster than the first one, because does not use look-up tables for XOR, but have the disadvantage that the encryption key requires preprocessing that is done on CPU. The maximum performance achieved on a GeForce 7900 GTX (255 GFlops), using the second method and 128-bit ECB AES is 108 Mbytes/sec (excluding the time required to prepare the key), and for the first method is 40 Mbytes/sec.

Other implementation was presented in the *"GPU Gems 3"* book [4], and it marks the transition to a new generation of GPUs with full support for integer operations inside the *shaders*. Implementation is based on OpenGL API, and the code for the *shaders* is written in assembly specific to GPUs series 8 from nVidia.

Yamanouchi [4] presents two methods for encryption on the new GPU generation: first method uses the vertex *shader* stage for encryption, the input data is stored in a Vertex Buffer and the result is written in a Vertex Buffer using the new feature of the Direct3D 10 class GPUs, output stream. The second method uses the pixel *shader* stage for encryption, the input data is stored in a texture and the output is also stored in another texture. The performance for 128-bit ECB AES obtained on GeForce 8800 GTS (403 GFlops) by the first method is 53 Mbytes/sec and for the second method is 95 Mbytes/sec. It should be noted that this implementation is the first for which there is no preprocessing done on input data or encryption key, so using this method we do not waste CPU time for preprocessing the data, the only functions performed by the CPU being the initialization of the GPU, uploading and downloading data from the GPU and sending draw commands to the graphics API. This implementation can be considered as the first implementation where the GPU is used as a cryptographic coprocessor, and have similar or higher performance than a single CPU. For example, O*penSSL* implementation reported a speed of 55 MBytes/sec on the same system.

With the advent of new Direct3D 10 compatible GPUs, GPU manufacturers have realized that this new generation has the opportunity to be used not only for graphics, but also for other applications that contain pieces of code very parallelizable. In this sense, they have created new tools and programming languages designed to allow the programmer to reach full power of GPU processing. nVidia created CUDA (Compute Unified Device Architecture), a C- based programming language with specific extensions for parallel execution of code. An implementation of AES using CUDA is shown by Manavsky [5], the author reporting speeds of up to 320 Mbytes/sec for 128-bit ECB AES on GeForce 8800 GTX (518 GFlops).

## 3. Direct3D 10 GPU Solution

To be able to implement general purpose algorithms using the GPUs we first need some means to program it, to be able to process the input data. On almost all the current video cards (those that are at least Direct3D 8 capable), the GPUs support *shaders* – small specific programs that are executed in parallel on multiple threads and their main purpose is to compute the shades of colors for the drawn objects (that's why they are called *shaders*). For Direct3D 10, there are three types of *shaders* that can be executed in parallel: *vertex shaders, geometry shaders* and *pixel shaders*. The *shaders* are executed once for each drawn element: vertex, geometry primitive and pixel. *Shaders* can read data from textures or constant buffers, but they can't write data directly. (In this implementation of AES algorithm, we don't use geometry *shaders*, so further information about those is not provided here – refer to [6], [7], [8] for more details.) All the values returned by the *vertex shader* are interpolated and passed as parameters to the *pixel shader*. The values returned by the *pixel shader* are passed to the *output assembly* stage of the pipeline and finally written in the render target texture. For implementation of cryptographic functions on the GPUs, we also need support for integer numbers inside the *shaders*, and therefore Direct3D 10 API is necessary, being the first version to provide support for integer operations.

To perform data encryption using AES algorithm, first we need to run the encryption code for each 16 bytes that make up a block, and because we need to read the input data exactly as found in memory, without other conversions applied to it, we will use as input a texture that contains only integer numbers. By coincidence, we can use the texture type *DXGI_FORMAT_R32G32B32A32_UINT*, which means that it will be used 32-bit per color component, resulting in a total of 128 bits (16 bytes) for each pixel from the texture. Thus using this format for both, input and output data, the encryption code will read an input pixel from the texture, apply on it the AES algorithm and write the result pixel into the output texture that has the same format as the input. This type of data processing, that is applied on a single pixel and has as a result another pixel, is very well suited for *pixel shader* stage defined in the Direct3D pipeline. The major advantages for running the encryption using *pixel shaders* are:

- Pixel shader code will represent exactly the encryption/decryption of a block of data;
- Because there are no dependencies between the processing blocks (pixels), every block can be processed in parallel;
- Results will be written automatically in the output texture by the pipeline stage output-assembly.

Normally, when Direct3D API draws something, the result is displayed in a viewport visible on the screen, but this is nit desired when using the GPU for other purposes, like encryption. To resolve this, we can create the device that performs all the drawing, without viewing area and the results will be stored in an off-screen texture created especially for this purpose (it's size – both vertical and horizontal, and format must match the size and format of the texture that holds the input data).

Since the creation of the resources (textures, vertex buffers and resource views) is an expensive operation, we will create all the resources needed for encryption and decryption when the application starts. This approach means that the input and output textures will have a predefined size, independent of data size that needs to be processed.

To process all the pixels in the input texture and produce an appropriate result for each pixel, the textures should be used so that no transformation are applied to the used coordinates, thus resulting in a one to one correspondence between pixels in the input and output textures. To achieve a one to one correspondence between input and output pixels, we need to draw a quad over the entire surface of the output texture (full screen aligned quad). To draw the full screen aligned quad we will use vertex coordinates in the screen space, and to be able to access the input texture we will use texture coordinates mapped in such a way that exists one to one correspondence between the input and output pixels. The vertex coordinates and texture coordinates used to draw the *full screen aligned quad* are presented in figure 1.

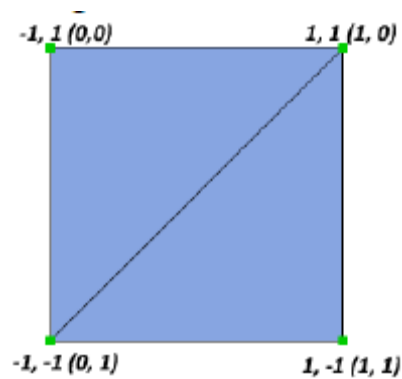

*Figure 1. Full screen aligned quad used to achieve one to one mapping between input and output pixels.*

Both the vertex *shaders* and pixel *shaders* are executed in parallel on a variable number of GPU cores, and their execution order is dependent on the used GPU, but we know for sure that the pixel *shader* will be executed for each pixel in the render target, and the received parameters are the return values of vertex *shader*, interpolated for each pixel. Because the input and output texture are created once with a predefined size, it is possible that in some cases we may not have enough data to fill the entire texture and we need to skip from processing the pixels that are not initialized, and for this we use *scissor* *rectangles* that describe the region of interest that needs to be drawn (see figure 2 for a visual representation).

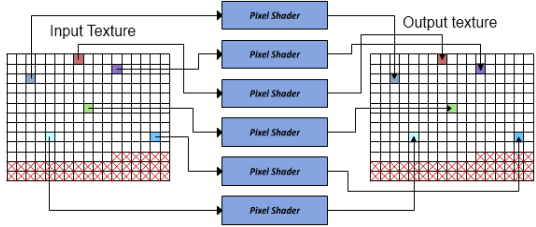

*Figure 2. One to one correspondence between input and output pixels*

Encryption key, along with the rest of the constants needed by the AES algorithm can be held in constant buffers or textures, but we must consider how these constants are accessed, because some GPU models does not have cache for constant buffers and they have cache only for textures. Precisely for this reason, the encryption key with the initial vector will be held in the constant buffers, because they will be accessed sequentially, and the remaining tables will be held in a texture, each table being held on a full line of the texture (all GPUs have caches of different sizes for texture accesses, making random accesses much faster, but sequential accesses on textures are slower than sequential accesses to constant buffers).

Since pixel *shaders* are run in parallel for multiple pixels at once, and there are no synchronization operations between threads, we can implement on the GPU only encryption modes that involve a parallelizable approach. That's why we can implement encryption and decryption for ECB and CTR mode, but for CBC mode only decryption can be implemented, because the encryption of a block depends on the result of the previous block.

For CBC mode, decryption is done entirely on the GPU, but only the first block is XOR-ed with the previous encrypted block on the CPU (this is done to avoid conditional expressions in the *shader* because they have a big impact on performance). To read the previous pixel in the input texture (previous encrypted block), texture coordinates are updated as following:

tx += (tx.x == 0)?int3(txSize.x - 1, -1, 0): int3(-1, 0, 0); (2)

For CTR mode, a counter value must be computed for each pixel (block), and this value must be added to the initial vector that is stored on 128-bits. The counter value is computed taking into account the texture coordinates then this value is added with the initial vector (a vector with 4 elements on 32-bits):

```
int   blockID  =   tx.x   +   tx.y   *
txSize.x;
state = IV;state.x += blockID;
if( state.x < IV.x ) {
  state.y ++;
  if( state.y == 0 ) {
    state.z ++;
    if( state.z == 0 )
        state.w ++;
  }
}
```

Once the drawing of the full screen aligned quad has ended (encryption/decryption has completed), we need to copy the results from the render target (output texture). Reading the output data is accomplished by mapping the texture memory (found in the VRAM) in the address space of the CPU and then performing a copy of the results in the system RAM. This mapping is accompanied by synchronization of the GPU, so that mapping occurs only after all draw operations are complete. Mapping of the input texture must also be done when we copy input data into the input texture. Because both input and output must be copied from/to RAM addressable by the CPU to RAM addressable by the GPU, we lose time with this operations because all data must be sent through the PCI Express bus which has a lower speed than the main memory bus. Copying data through the PCI Express bus is made more effective as the amount of data being transmitted is greater and for out implementation the size of the data is limited by the input/output texture size.

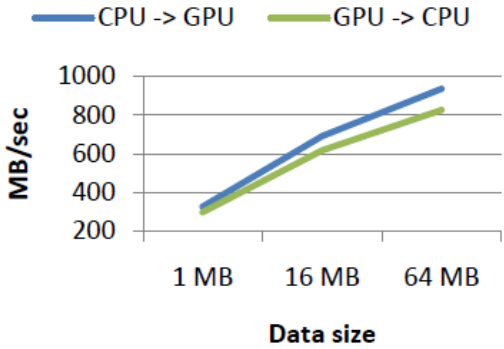

*Figure 3. The bandwidth between CPU and GPU, taking into account the size of the transferred data.*

According to Direct3D 10 API, the minimum texture size is 8192x8192 pixels and is guaranteed that textures smaller than 128 MB can be created and placed inside VRAM for optimal speed. Because of this and the fact that Windows Vista and 7 implements a watchdog timer that resets the GPU when a draw operation takes longer than 2 seconds (it is assumed that is something wrong with the GPU or the *shader* that is running), we choose the size of the input and output textures of 64 MB (2048x2048 with 128 bits per pixel) representing a compromise between execution speed and the proper execution of the application.

## 3.1 Performance analysis

In the figure 4 we have the results for 128- bit and 256-bit ECB AES on a system with Intel Core 2 Quad Q9450 CPU @2.66 GHz (42.5 GFlops) and nVidia GeForce 9800GTX (648 GFlops). The encryption is measured on one, two, three and four CPU cores and on the GPU for different input data size. The CPU implementation is based on the same algorithm that runs on GPU, it was compiled for x64 architecture to take advantage of extra registers and is comparable with the implementation from *OpenSSL* library [9].

For the GPU implementation there is a major overhead for cases where data is generally smaller than 64K. This overhead is mainly due to data copy from the RAM addressable by the CPU in the RAM addressable by the GPU and vice versa (see Figure 3). Other major overhead is sending the drawing commands to the video card driver, however as can be seen from the figure 4, for larger data size middle- class GPUs are performing faster than two cores on a high end CPU.

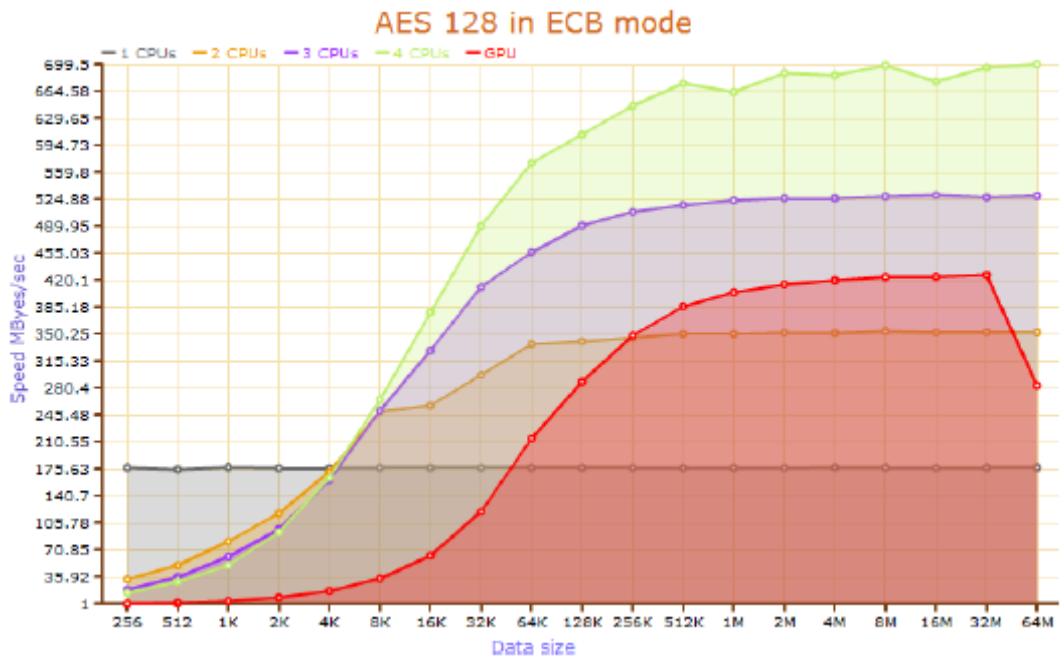

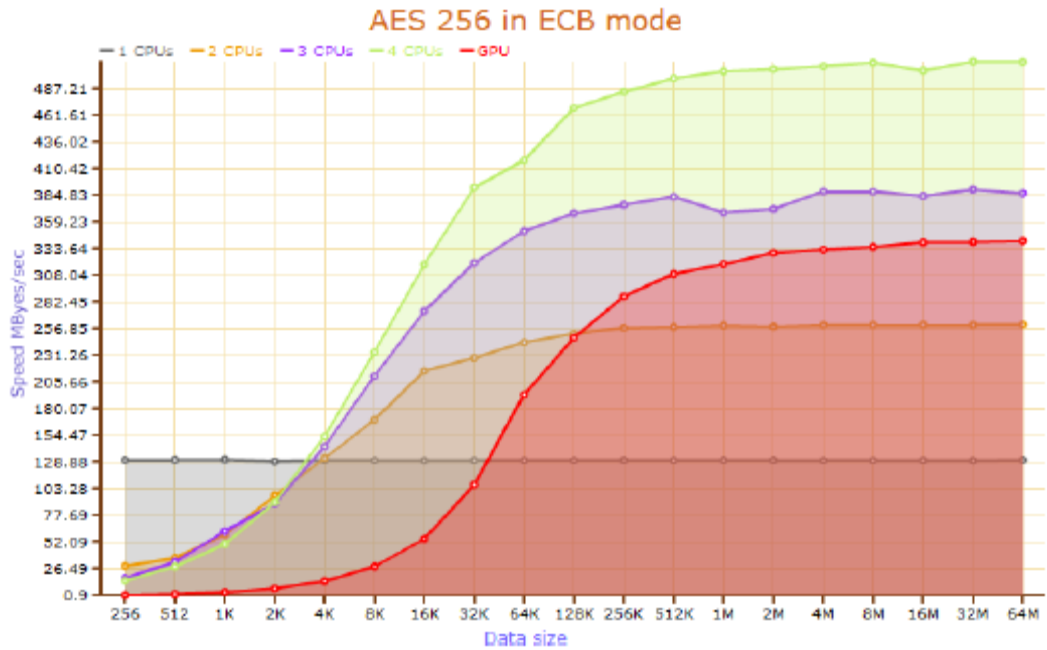

*Figure 4.Speed results for AES in ECB mode running on the GPU and CPU*

## 4. Conclusions

Because data security is becoming increasingly important and most current computers have GPUs compatible with Direct3D 10 API, operations like encryption and decryption can be performed on GPU (which most of the time is not used), so freeing up processing resources from the CPU that can be used for other tasks. Thus, using the GPU as a cryptographic accelerator, we can perform operations like backup with encryption without the use of CPU resources. The GPU is strong enough for such operations even on laptops that are usually equipped with low-level graphics cards in terms of graphics performance.

Using the GPU in applications that work with small data, such as encryption of communications between server and client is not currently recommended because each operation of encryption/decryption have a big overhead compared to the same operation performed on the CPU, this overhead is largely due to copy operations of data between RAM addressable by the CPU and RAM addressable by the GPU.